\documentclass[conference]{IEEEtran}
\IEEEoverridecommandlockouts
\usepackage{color}
\usepackage{multirow}
\usepackage{booktabs}
\usepackage{url,times,booktabs,tabularx,amsfonts,siunitx}
\usepackage{algorithmic}
\usepackage{setspace}
\usepackage{graphicx}
\usepackage{textcomp}
\usepackage{xcolor}
\usepackage{latexsym}
\usepackage{bbding}
\usepackage{footnote}
\usepackage{arydshln} % 导入绘制虚线的宏包
\usepackage{hyperref}

\usepackage{cite}
\usepackage{amsmath,amssymb,amsfonts}
\usepackage{algorithmic}
\usepackage{graphicx}
\usepackage{textcomp}
\usepackage{xcolor}

\def\BibTeX{{\rm B\kern-.05em{\sc i\kern-.025em b}\kern-.08em
    T\kern-.1667em\lower.7ex\hbox{E}\kern-.125emX}}

\title{FlowSep: Language-Queried Sound Separation with Rectified Flow Matching}

\author{\IEEEauthorblockN{Yi Yuan, Xubo Liu, Haohe Liu, Mark D. Plumbley, Wenwu Wang}
\IEEEauthorblockA{\textit{Centre for Vision, Speech and Signal Processing~(CVSSP)} \\
\textit{University of Surrey,
Guildford, UK} 
}
}
\begin{document}

\maketitle

\begin{abstract}

Language-queried audio source separation (LASS) focuses on separating sounds using textual descriptions of the desired sources. Current methods mainly use discriminative approaches, such as time-frequency masking, to separate target sounds and minimize interference from other sources. However, these models face challenges when separating overlapping soundtracks, which may lead to artifacts such as spectral holes or incomplete separation.
Rectified flow matching~(RFM), a generative model that establishes linear relations between the distribution of data and noise, offers superior theoretical properties and simplicity, but has not yet been explored in sound separation. In this work, we introduce FlowSep, a new generative model based on RFM for LASS tasks. FlowSep learns linear flow trajectories from noise to target source features within the variational autoencoder (VAE) latent space. During inference, the RFM-generated latent features are reconstructed into a mel-spectrogram via the pre-trained VAE decoder, followed by a pre-trained vocoder to synthesize the waveform. Trained on $1,680$ hours of audio data, FlowSep outperforms the state-of-the-art models across multiple benchmarks, as evaluated with subjective and objective metrics. Additionally, our results show that FlowSep surpasses a diffusion-based LASS model in both separation quality and inference efficiency, highlighting its strong potential for audio source separation tasks. Code, pre-trained models and demos can be found at: \url{https://audio-agi.github.io/FlowSep_demo/}.

\end{abstract}

\begin{IEEEkeywords}
Language-queried audio source separation, sound separation, rectified flow matching, multimodal learning
\end{IEEEkeywords}

\section{Introduction}
\label{sec:intro}
Audio source separation systems aim to separate specific sound sources from audio mixtures. Previous research has made significant progress in various domains, including speech~\cite{speech1,speech2}, music~\cite{resunet,music2} and acoustic events~\cite{universal2,liu2024audio}. Recently, there has been growing interest in separating target audio sources using natural language queries as known as the Language-Queried Audio Source Separation~(LASS) ~\cite{lassnet} task. 
% Unlike previous label-queried methods~\cite{query1,query2}, taking natural language as inputs addresses the limitations of predefined sound categories, enabling flexible open-domain source separation. 
LASS provides a useful tool for future source separation systems, allowing users to extract audio sources via natural language instructions. These systems could be useful in many applications, such as automatic audio editing~\cite{audioedit,audioedit2}, multimedia content retrieval~\cite{clip}, and audio augmented listening~\cite{augmentlisten}.

The first attempt is the LASS-Net~\cite{lassnet}, which employs a BERT~\cite{bert} network to encode textual queries, a ResUNet~\cite{resunet} module to predict the spectrogram masks of the target source. AudioSep~\cite{audiosep} leverages contrastive multimodal pretraining models (e.g., CLAP~\cite{clap}) and scales up the training data to \num{14,000} hours, which achieves state-of-the-art results and shows promising zero-shot separation performance.
However, both LASS-Net~\cite{lassnet} and AudioSep~\cite{audiosep} primarily use masking-based discriminative methods (e.g., spectrogram masking \cite{speech2}) to separate target audio sources from mixtures. These systems may encounter challenges when dealing with overlapping sound events~\cite{tse_ref}. In particular, the masks generated by these models may be excessive or insufficiently selective, which leads to artifacts such as spectral holes or incomplete separation~\cite{dpm-tse}. These limitations affect its effectiveness in real-world scenarios with diverse and dynamic acoustic environments. 

Recently, researchers have been exploring separation source systems using non-discriminative models, such as generative models including Generative Adversarial Network (GANs)~\cite{chen2024mdx} and diffusion-based models ~\cite{dpm-tse}. Due to their generative nature, these models have the potential to enhance the perceptual quality of the separated sources and improve the overall subjective quality. Rectified Flow Matching (RFM) ~\cite{flow_matching} is a recently proposed generative model. Similar to diffusion-based systems that learn to gradually denoise the output from general distribution, RFM models linear the relationships between data distributions and noise and provide superior theoretical properties and simplicity. However, it has not yet been explored in sound separation tasks.

% Liu et al.~\cite{speech_generation} first developed a high-quality speech separation system based on generation models with convolutional neural network (CNN)-based encoder. Hai et al.~\cite{dpm-tse} presented a label-based target sound extraction~(TSE) method with a generative model i.e. diffusion probabilistic modelling~(DPM).

\begin{figure*}[htbp]
    \centering
    \includegraphics[width=0.8\linewidth]{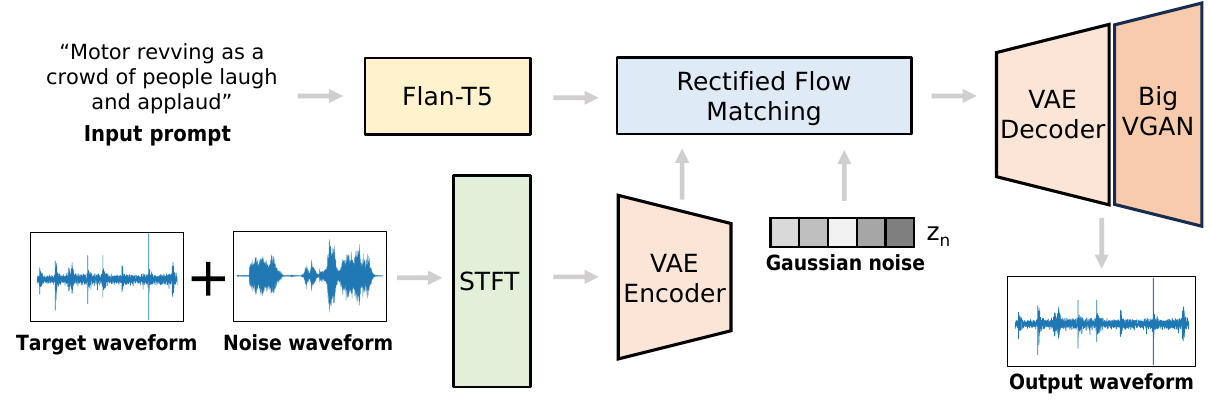}
    \caption{The architecture of FlowSep. FlowSep consists of four main components: (1) a FLAN-T5 encoder for text embedding; (2) a VAE for encoding and decoding mel-spectrograms; (3) an RFM module for generating audio features within the VAE latent space; (4) a BigVGAN vocoder to generate the waveform.}
    \label{fig:overview}
\vspace{-0.3cm} % 减少0.5厘米的垂直空间
\end{figure*}

In this work, we propose FlowSep, a LASS system using rectified flow matching. FlowSep employs a FLAN-T5 encoder~\cite{t5} to embed the textual queries, followed by an RFM-based separation network. Specifically, FlowSep learns linear flow trajectories from noise to target features within the VAE latent space, conditioned on the query embeddings. During inference, the RFM-generated latent features are decoded into a mel-spectrogram using a pre-trained VAE decoder and then converted to a waveform with a pre-trained vocoder.
% Unlike previous discriminative approaches, the audio sources extracted by the generative models may not be well aligned with the waveform of the original source in timing and amplitude. Consequently, traditional methods widely used for evaluating audio separation may not be appropriate as shown in~\cite{speech_generation}. Thus, we primarily use generative-based objective and subjective metrics to evaluate the performance of FlowSep.
We train FlowSep using \num{1,680} hours of audio data including AudioCaps~\cite{audiocaps}, VGGSound~\cite{vggish} and WavCaps~\cite{wavcaps}. Experimental results across multiple datasets demonstrate that FlowSep significantly outperforms previous state-of-the-art models, such as AudioSep \cite{audiosep} in both objective and subjective metrics, showing promising separation performance in real-world scenarios. Additionally, comparisons between FlowSep and a diffusion-based LASS model indicate that RFM enhances both the output quality and inference efficiency, highlighting its strong potential for sound separation tasks.

This paper is organized as follows. Section~\ref{sec:system} introduces the proposed system, followed by the dataset and evaluation methods applied to evaluate the performance of FlowSep in Section~\ref{sec:dataset}. Section \ref{sec:exp} presents the experimental results and conclusions are given in Section~\ref{sec:conclusion}.

% P1: [audio source separation, then introduction of LASS and its application]

% p2: Recent methods for LASS task; [First model, LASS-Net] + [several works addressing the lack of training data issue by using multimodal learning e.g., CLIPSep] + [AudioSep] (borrow texts from AudioSep paper)

% p3: [Existing LASS work use discriminative masking based method (e.g., time-frequency masking), then what's its limitation, may ask ChatGPT xD][Analysis of why generative modelling is potential for LASS] [Some introductions of generative speech separation/enhancement and sound separation work, while these methods have not explored yet in LASS]. (check this paper and its reference for some useful texts: https://arxiv.org/pdf/2310.04567)

% p4: In this work, we propose:

\section{Method}
\label{sec:system} 
We propose FlowSep, an RFM-based generative model for language-queried audio source separation. FlowSep consists of four main components: a FLAN-T5 encoder for text embedding, a VAE for encoding and decoding mel-spectrograms, an RFM-based latent feature generator for predicting audio features within the VAE latent space, and a GAN-based vocoder~\cite{bigvgan} to generate the waveform. The model architecture and the separation workflow are illustrated in Figure~\ref{fig:overview}.

\subsection{Text Encoder}
Unlike AudioSep \cite{audiosep}, which uses a contrastive language-audio pre-training (CLAP) encoder~\cite{clap} for text query embedding, we use a FLAN-T5 encoder~\cite{t5}. As motivated by advancements in audio synthesis ~\cite{reaudioldm,tango2}, where FLAN-T5 has demonstrated better performance compared to CLAP.

\subsection{Latent Feature Generator}
The latent feature generator is a UNet-based~\cite{unet, audioldm, audioldm2} network with cross-attention modules to process the T5-embedding. We first introduce the RFM approach, followed by a description of the channel-conditioned generation approach, where we adapt RFM to the LASS task.

% \subsubsection{Latent Diffusion Model}
% \label{sec:diffusion}
% The general diffusion approach involves two processes: a forward process that gradually adds Gaussian-based noise into the target vectors and a reverse process that predicts the noise of the vector added in during forward steps. During the forward stage, the presentation of the feature in latent space $z_0$ is transformed into a standard Gaussian distribution $z_n$ through a noise injection procedure $q(\cdot)$ as :
% \begin{equation}
% q(\boldsymbol{z}_{n}|\boldsymbol{z}_{n-1})=\mathcal{N}(\boldsymbol{z}_{n};\sqrt{1-\beta_{i}}\boldsymbol{z}_{n-1},\beta_{i}\boldsymbol{I})
% \end{equation}
% \noindent
% where $1 -\beta_{n}$ controls the noise level of the Gaussian noise added within each step and $n$ is the maximum forward step. 
% Then during the reverse stage, LDM learns to estimate the distribution of noise  $\boldsymbol{\epsilon}_{\theta}$ in step $z_i$ to reconstruct the vector for step $z_i-1$, illustrate the loss function~\cite{DDPM} as:
% \begin{equation}
% \label{trainingobjective}
% L_{D}(\theta)=\mathbb{E}_{z_{n},\boldsymbol{\epsilon},i}\left \| \boldsymbol{\epsilon} - \boldsymbol{\epsilon}_{\theta}(z_{n},n,\boldsymbol{E}^{t}) \right\|^2_{2}
% \end{equation}
% where $\epsilon$ denotes the Gaussian noise on step $i$. Here we take the text embedding $\boldsymbol{E}^{t}$, the time step $\textit{i}$ as the condition. 

\subsubsection{Rectified Flow Matching}

RFM aims to predict the vector field $\mu$ which maps a linear pathway between the noise distribution $z_0 \in N[0,I]$ and target feature $z_1$. Specifically, RFM learns this UNet neural network $\mu~(\cdot, \theta)$ to predict the optimal transformation path for any noisy audio feature $z_t$ with a random flow step $t \in[0,1]$ and parameter $\theta$. During training, the RFM first computes the noisy version of each data $z_1$ as:
\begin{equation}
z_{t} = (1 - (1-\sigma)t)z_{0} + tz_{1}
\end{equation}

\noindent where $\sigma$ is a relatively small value~\cite{audiobox} and chosen empirically as $1\times 10^{-5}$ in FlowSep. Taking both mixture audio feature $z^{m}$ and query embedding $\boldsymbol{E}$ as condition, the target of RFM is summarized as:
\begin{equation}
v = z_1 - (1-\sigma)z_{0}
\end{equation}
\begin{equation}
L_{\text{RFM}}(\theta) = \mathbb{E}_{t, z_1,z_0} \left\| \mu(z_t,\boldsymbol{E},\boldsymbol{z}^{m}) - v\right\|^2
\end{equation}
Then during the inference stage, the output latent vector $\hat{z}_1$ is sampled from a noise $z_n$ sampled from Gaussian distribution. By leveraging the theoretical properties and simplicity of RFM (e.g., linear flow trajectories), the inference process can be completed in a relatively small number of steps (e.g., fewer than $10$ steps), greatly improving the system's efficiency.

\begin{figure}[!tbp]
  \centering
  \includegraphics[width=\linewidth]{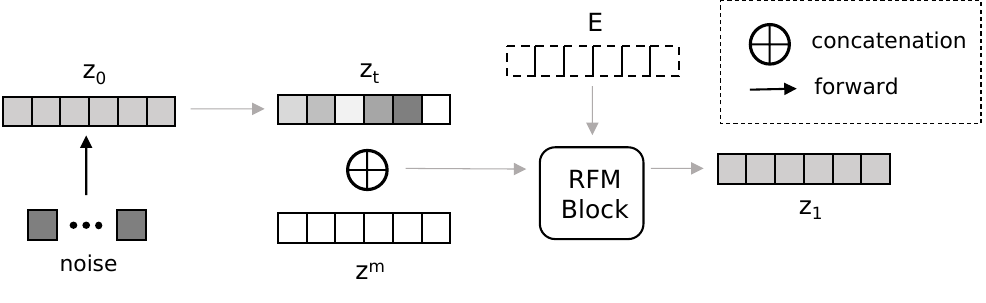}
  \caption{The channel-concatenation conditioning mechanism}
  \label{fig:two_channel}
  \vspace{-0.2cm} % 减少0.5厘米的垂直空间
\end{figure}

\subsubsection{Channel-Conditioned Generation}
FlowSep aims to generate the target feature $\hat{z}_1$ based on the mixture waveform and text query. Hence, we propose channel-conditioned generation to guide the model by taking the mixed audio as input channel conditions. As shown in Figure~\ref{fig:two_channel}, both standard Gaussian distribution $z_t$ and the latent vectors of mixture mel-spectrogram $z^{m}$ are concatenated into the input channel before being forwarded into the RFM module. In this way, the additional condition, $z^{m}$, is considered as extra information within the input so the feature can be processed in parallel with the target latent vector by the RFM model. The noise adding forward process does not affect the mixture channel. 

\subsection{VAE Decoder and GAN Vocoder}
FlowSep leverages a combination of a VAE and a GAN network for reconstructing the target waveform. The VAE encodes the mel-spectrogram into an intermediate representation~(latent space feature), which is then decoded back to its original form. Following this, a GAN-based vocoder is trained to convert the decoded output into the waveform. In FlowSep, we use the state-of-the-art universal vocoder BigVGAN~\cite{bigvgan}.

\begin{table*}[htbp]
\caption{Objective evaluation on LASS, where AC, VGG and ESC are short for AudioCaps~\cite{audiocaps}, VGGSound~\cite{vggish} and ESC50~\cite{esc-50} respectively.FAD does not apply on unprocessed data as it calculate the difference between two groups of audio, while the target and mixed audio share the same audio events.} 
\centering
\small
\resizebox{0.95\textwidth}{!}{%
\begin{tabular}{ccccc|ccccc|cccc}
\toprule
                     \multirow{3}{*}{Model}  & \multicolumn{4}{c}{FAD $\downarrow$} &  \multicolumn{5}{c}{CLAPScore $\uparrow$}& \multicolumn{4}{c}{CLAPScore$_{A}$ $\uparrow$}\\
 \cmidrule(lr){2-14}
&  AC & DE-S  & VGG & ESC  &  AC & DE-S  & DE-R& VGG & ESC  &  AC & DE-S  &  VGG & ESC \\
\midrule
Unprocessed
&    $-$    &  $-$  &  $-$   & $-$ & $11.9$ &$23.2$ &  $22.7$&    $13.6$    &  $19.1$   &  $64.9$   & $71.3$  &   $66.7$ &  $71.3$   \\
LASS-Net~\cite{lassnet}
&    $5.09$    &  $1.83$  &  $3.09$   & $3.28$ & $14.4$ &$24.4$ &  $25.3$ &    $17.4$    &  $20.5$   &  $70.2$   & $76.6$ &   $69.5$ &  $79.6$   \\
AudioSep~\cite{audiosep}
 &    $4.38$    &  $1.21$ & $2.30$     & $1.93$   & $13.6$ &$26.1$ &  $29.7$ &    $19.0$    &  $21.2$   &  $69.6$   & $78.9$ &  $72.4$ &  $80.5$   \\
FlowSep
&  $\mathbf{2.86}$   &  $\mathbf{0.90}$  &   $\mathbf{2.06}$     & $\mathbf{1.49}$   & $\mathbf{21.9}$ &$\mathbf{26.9}$&  $\mathbf{31.3}$ &    $\mathbf{19.5}$    &  $\mathbf{22.7}$   &  $\mathbf{81.7}$   & $\mathbf{80.1}$  &  $\mathbf{73.2}$ &  $\mathbf{80.7}$ \\
\bottomrule
\end{tabular}
}
\label{tab:subjective}
\vspace{-0.4cm} % 减少0.5厘米的垂直空间
\end{table*}

\section{Dataset and Evaluation metrics}
\label{sec:dataset} 

% This section describe the datasets used for training and testing FlowSep, along with the metrics for evaluation.

\subsection{Dataset}
\label{sec:database}
\subsubsection{Training Set}
we use $1,680$ hours of audio from AudioCaps~\cite{audiocaps}, VGGSound~\cite{vggish} and WavCaps~\cite{wavcaps} for training.  \\
\noindent\textbf{AudioCaps~\cite{audiocaps}} is the largest publicly available audio dataset consisting of $10$-second audio clips paired with human-annotated captions. As the subset of AudioSet~\cite{audioset}, AudioCaps contains $49,837$ training clips and $957$ testing samples. \\
\noindent \textbf{VGGSound~\cite{vggish}} is a large-scale audio dataset consisting of approximately $200,000$ audio clips sourced from YouTube. Each audio clip in VGGSound has a duration of $10$ seconds, while the audio clips in VGGSound are annotated with a set of $309$ audio event textual labels, rather than detailed captions. \\
\noindent \textbf{WavCaps~\cite{wavcaps}} is an audio dataset with machine captions generated using Large Language Models~(LLM). We used audio samples shorter than $10$ seconds and selected a total of \num{400,000} audio clips for training. 
\subsubsection{Test Set}
We evaluate performance using five different benchmarks: the VGGSound and ESC50~\cite{esc-50} test datasets, as applied in the evaluation benchmark proposed by AudioSep \cite{audiosep}, the AudioCaps testing set, and the two official evaluation datasets from DCASE2024 Task 9\footnote{https://dcase.community/challenge2024/task-language-queried-audio-source-separation}. During evaluation, it is ensured that target and noise sources in each mixture do not share any overlapping sound classes.

\noindent \textbf{VGGSound} testing set selected $200$ clean and distinct audio samples as the target audio and $10$ audio samples from the remaining testing set. The testing set consists $2,000$ mixtures by mixing each target sample and noise sample with random LUFS loudness between $-35$ and $-25$ dB. \\
\noindent \textbf{ESC50 ~\cite{esc-50}} evaluation set provides a total of $2,000$ mixtures, where each clip is mixed with a sample from the ESC50 with a signal-to-noise ratio~(SNR) at $0$ dB. \\
\noindent\textbf{AudioCaps} testing set consisting of $928$ samples, where we take each audio clip as the target source and mix it with a noise source selected from the testing set under random SNR rate between $-15$ and $15$ dB. We select the first caption of the target source as the query for separation.  \\ 
\noindent \textbf{DCASE2024 Task 9} 
provides two evaluation sets which can be directly applied for evaluation: DCASE-Synth (DE-S) and DCASE-Real (DE-R). In detail, DE-S includes $3,000$ synthetic mixtures mixed from $1,000$ audio clips with an SNR rate ranging from $-15$ to $15$ dB. DE-R consists of $100$ audio clips collected from real-world scenarios. Each audio clip contains at least two overlapping sound sources. Each audio clip was annotated their component sources using text descriptions, so that each clip can be used as a mixture from which to extract one or more of the component sources based on a text query. Each audio clip in DE-R was labeled with two such text queries.

\subsection{Evaluation Metrics}
Unlike discriminative networks used in previous systems~\cite{audiosep}, FlowSep does not separate audio events by masking the mixture input. The separated results are not strictly aligned with the target audio samples in the temporal dimension. Hence, traditional sample-level objective metrics for source separation tasks, such as source-to-distortion ratio (SDR), are not well-suited for evaluating the proposed system. Instead, we assessed its performance using five different metrics. For objective evaluation, we first use frechet audio distance (FAD)\cite{fad}, a common metric for evaluating generation systems. Next, we apply two metrics based on CLAP~\cite{clap} including CLAPScore~\cite{clapscore}, a reference-free metric measuring similarity between the output audio and the text query, and CLAPScore$_{A}$, which evaluates the similarity between the output audio and the target audio. For human evaluation, we follow the official methods used in the DCASE2024 Task 9 Challenge, which include assessing the relevance between the target audio and the language query (REL) and the overall sound quality (OVL). Both OVL and REL metrics are rated on a Likert scale from $1$ to $5$. All five datasets were evaluated using objective metrics, and we further conducted subjective evaluations on the AudioCaps, DE-S, and DE-R datasets, using a subset of 50 samples randomly selected from each. The subjective evaluations were performed on ten different listeners, with six researchers in audio and speech and four from other fields.

\begin{figure*}[htbp]
    \centering
    \includegraphics[width=1.0\linewidth]{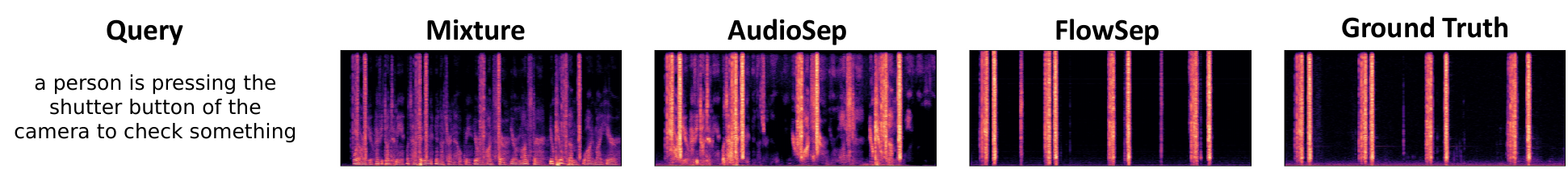}
    \caption{A case study of separation results on DE-S test set, as compared with the ground truth. More results can be found online from \href{https://audio-agi.github.io/FlowSep_demo/}{online}.}
    \label{fig:case}
\vspace{-0.3cm} % 减少0.5厘米的垂直空间
\end{figure*}

\section{Experiments and Results}
\label{sec:exp} 

\subsection{Data Processing}
For training data processing, we first apply the PANNs~\cite{kong2020panns}, an off-the-shelf audio tagging model, to label the audio clips to ensure that every two audio clips used for creating synthetic mixture do not share any overlapping sound source classes. Then, the audio segments are padded or cropped to $10$ seconds with $16$ kHz sampling rate. We create synthetic training mixtures with a random SNR between $-15$ and $15$ dB. The mixed waveform is then calculated through STFT under a frame of $1024$ and a hop size of $160$ to obtain the mel-spectrogram. For the language query, all the textual captions are converted into lower cases and punctuation is removed. 

\subsection{Experimental Details}
We first train a $16$ kHz BigVGAN \cite{bigvgan} on the training datasets. For the encoder and decoder of the FlowSep, we apply the FLAN-T5-large~\cite{t5} checkpoint and the pre-trained VAE model from AudioLDM~\cite{audioldm} model. We freeze all the other components and the RFM model is trained for $1$M steps with a batch size of $8$ and a base learning rate of $5 \times 10^{-5}$. We use the publicly released LASS-Net \cite{lassnet} and AudioSep \cite{audiosep} models as baselines. For the diffusion-based network, denoted as DiffusionSep, we replace the RFM with general diffusion-based loss \cite{audioldm} and train the system using the same parameters. 

% \begin{figure*}[htbp]
%     \centering
%     \includegraphics[width=0.8\linewidth]{case.pdf}
%     \caption{Visualization of separation results on DE-R}
%     \label{fig:case}
% \vspace{-0.3cm} % 减少0.5厘米的垂直空间
% \end{figure*}

\subsection{Experimental Results}

% For both objective and subjective evaluations, we compare the baseline models with FlowSep with a default of $10$ steps through the ODE solver. We evaluate both the baseline systems and FlowSep across five evaluation datasets. Except for AudioCaps, the other evaluation datasets are zero-shot for FlowSep. Experimental results are shown in Table~\ref{tab:objective} and Table~\ref{tab:subjective}. FlowSep outperforms the baseline models by a significant margin on all eval datasets. The results of AudioCaps testing sets, which contain complex and overlapping audio events, highlight the ability of FlowSep on separating overlapped soundtracks. The results on DCASE2024-Real show an enhanced capability of the proposed system in real-world cases. The proposed system achieves better performance on ESC50 and DCASE2024 testing sets across all the metrics, illustrating an enhanced capability on zero-shot cases than the SoTA models. In addition, evaluations on subjective metrics~(REL and OVL) illustrate that FlowSep not only generates relevant audio events based on natural language descriptions but also presents audio with better overall quality. Furthermore, FlowSep achieves a better CLAPScore$_{A}$, which presents the similarity between target audio and separated output. With an average CLAPScore$_{A}$ of $79$ across the four evaluation sets, FlowSep effectively identifies and extracts relevant features, resulting in better alignment and more correct content with the ground truth.
We evaluated FlowSep against baseline models using $10$ steps with the ODE solver across five datasets, with the AudioCaps and VGGSound testing set being the non-zero-shot test dataset for FlowSep. Results in Table~\ref{tab:objective} and Table~\ref{tab:subjective} show that FlowSep significantly outperforms the baselines across all datasets. Results on DE-R highlight its enhanced capability in real-world scenarios, results achieved on ESC50 and two DCASE2024 testing sets show its effectiveness in zero-shot cases compared to state-of-the-art models. FlowSep also achieved a higher average CLAPScore$_{A}$ of 79 across four evaluation sets, showing better alignment and content accuracy with the ground truth. Furthermore, subjective metrics REL and OVL illustrate that FlowSep not only generates relevant audio events from natural language descriptions but also delivers higher overall perceptual quality.
% All evaluation clips are composed solely of audio samples from the testing database, ensuring that the model has no prior exposure to any information during training~(target signals and noise sources). From both results in Table~\ref{tab:objective} and Table~\ref{tab:subjective}, FlowSep outperforms the baseline models by a significant margin. The results of AudioCaps testing sets, which contain complex and overlapping audio events, highlight the ability of FlowSep on separating overlapped soundtracks. The results on DCASE2024-Real show an enhanced capability of the proposed system in real-world cases. The proposed system achieves better performance on ESC50 and DCASE2024 testing sets across all the metrics, illustrating an enhanced capability on zero-shot cases than the SoTA models. In addition, evaluations on subjective metrics~(REL and OVL) illustrate that FlowSep not only generates relevant audio events based on natural language descriptions but also presents audio with better overall quality. Furthermore, FlowSep achieves a better CLAPScore$_{A}$, which presents the similarity between target audio and separated output. With an average CLAPScore$_{A}$ of $79$ across the four evaluation sets, FlowSep effectively identifies and extracts relevant features, resulting in better alignment and more correct content with the ground truth.

\begin{table}[tbp]
\caption{Subjective evaluation results on LASS, where AC, DE-S and DE-R are short for AudioCaps~\cite{audiocaps}, DCASE-Synth and DCASE-Real respectively.} 
\centering
\small
\resizebox{0.48\textwidth}{!}{%
\begin{tabular}{cccc|ccc}
\toprule
                     \multirow{3}{*}{Model} & \multicolumn{3}{c}{REL$\uparrow$} & \multicolumn{3}{c}{OVL$\uparrow$}\\
 \cmidrule(lr){2-7}
&  AC &  DE-S & DE-R &  AC &  DE-S & DE-R\\
\midrule
% Mixture
% & $2.72$ & $2.96$ & $2.96$ & $2.16$ & $2.84$ & $2.96$\\
LASS-Net~\cite{lassnet}
&  $3.12$ & $2.96$ & $3.59$ & $2.16$ & $2.84$ & $3.88$\\
AudioSep~\cite{audiosep}
& $3.66$ & $3.24$ & $3.93$ & $2.69$ & $3.53$ & $4.02$\\
FlowSep
&  $\mathbf{4.08}$ & $\mathbf{3.62}$ & $\mathbf{4.11}$ & $\mathbf{3.98}$ & $\mathbf{3.72}$ & $\mathbf{4.26}$\\
\bottomrule
\end{tabular}
}
\label{tab:objective}
\end{table}

\subsection{Case Studies}
We visualized the separated spectrograms of AudioSep and FlowSep using an example from the DE-S testing set. As shown in Figure~\ref{fig:case}, the spectrogram generated by FlowSep is closely matching the ground truth. In contrast, AudioSep's results show incomplete separation with noticeable spectral gaps. The results highlight that, unlike previous discriminative approaches, FlowSep, as a generative network, demonstrates promising capabilities in source separation tasks. More case studies can be found online\footnote{https://audio-agi.github.io/FlowSep\_demo}.

\subsection{Efficiency Analysis}
We conduct an efficiency analysis of FlowSep in comparison with AudioSep and a diffusion-based LASS model we implemented, with the performance results presented in Table~\ref{tab:diffusion}. The inference experiments were conducted on the AudioCaps dataset using a single A100 GPU with 80 GB of memory. As compared with AudioSep, our generative approach does not present better efficiency than discriminative approaches, while FlowSep generates results with better quality. Furthermore, FlowSep surpasses the diffusion-based LASS model in both FAD and CLAPScore metrics with the same number of inference steps (e.g., $200$ steps). FlowSep-RFM maintains stable performance when the inference steps are reduced (e.g., only minor degradation when reducing the steps from $200$ to $100$) and achieves acceptable performance with even fewer steps (e.g., 10 steps). These findings underscore that RFM outperforms diffusion-based methods in separation tasks, and our proposed system can effectively reducing inference time while enhancing separation performance.

\begin{table}[tbp]
\caption{The efficiency analysis of FlowSep as compared with the baseline models. The VAE decoder and vocoder inference time is shown as superscripts. }
\centering
\small
\resizebox{0.48\textwidth}{!}{%
\begin{tabular}{ccccc}
\toprule

 \multirow{1}{*}{\textbf{Model}} &\multicolumn{1}{c}{\textbf{Infer-step}}  &\multicolumn{1}{c}{\textbf{Time(s)}} 
  & \multicolumn{1}{c}{\textbf{FAD}$\downarrow$} & \multicolumn{1}{c}{\textbf{CLAPScore} $\uparrow$}  \\
\midrule
AudioSep & $ - $&$ 0.06$ & $4.38$ & $13.6$  \\
\midrule
DiffusionSep & $50$&$4.9_{+0.12}$ & $4.52$ & $10.4$  \\

DiffusionSep & $100$&$9.4_{+0.12}$ & $3.46$ & $12.3$  \\

DiffusionSep & $200$&$18.1_{+0.12}$ & $2.76$ & $18.8$  \\
\midrule
FlowSep & $10$& $0.58_{+0.12}$ & $2.86$ & $21.9$  \\

FlowSep & $100$& $5.1_{+0.12}$ & $2.75$ & $22.8$  \\

FlowSep & $200$& $9.0_{+0.12}$ & $2.74$ & $23.1$  \\
% \midrule
\bottomrule
\end{tabular}

}

\label{tab:diffusion}
\vspace{-0.4cm} % 减少0.5厘米的垂直空间
\end{table}

\section{Conclusion}
\label{sec:conclusion} 
We introduced FlowSep, an RFM-based generative model for LASS, designed to leverage the strengths of RFM while overcoming the limitations of previous discriminative models for source separation. Experiments across various datasets demonstrate that FlowSep outperforms baseline models in both objective and subjective quality assessments. Furthermore, our results show that FlowSep present better performance on separation tasks than the diffusion-based model in both quality and the efficiency of inference. These findings highlight the significant potential of RFM for source separation tasks.
% In the future, we plan to explore the potential of FlowSep across a broader range of sound sources, including speech and music, or even semantic-guided queries,e.g, sounds that make people happy.

\section{ACKNOWLEDGMENT}
\label{sec:ack}
This research was partly supported by a research scholarship from the China Scholarship Council~(CSC), funded by British Broadcasting Corporation Research and Development~(BBC R\&D), Engineering and Physical Sciences Research Council~(EPSRC) Grant EP/T019751/1 ``AI for Sound'', a Research Gift from Google, and a PhD scholarship from the Centre for Vision, Speech and Signal Processing~(CVSSP), University of Surrey. 
For the purpose of open access, the authors have applied a Creative Commons Attribution~(CC BY) license to any Author Accepted Manuscript version arising.
% -------------------------------------------------------------------------
\bibliographystyle{IEEEtran}
% \bibliography{refs}


\end{document}